\documentclass{Interspeech}

\interspeechcameraready

\title{EME-TTS: Unlocking the Emphasis and Emotion Link in Speech Synthesis}

\author[affiliation={1}]{Haoxun}{Li}
\author[affiliation={1,*}]{Leyuan}{Qu}
\author[affiliation={1}]{Jiaxi}{Hu}
\author[affiliation={1,*}]{Taihao}{Li}

\affiliation{Hangzhou Institute for Advanced Study}{University of Chinese Academy of Sciences}{China}
\email{lihaoxun23@mails.ucas.ac.cn, leyuan.qu@ucas.ac.cn, hujiaxi23@mails.ucas.ac.cn,lith@ucas.ac.cn
}

\usepackage{comment}
\usepackage{float}  
\usepackage{subcaption} 
\usepackage{graphicx} 
\usepackage{adjustbox} 
\usepackage{hyperref}
\hypersetup{
    colorlinks=true, 
    urlcolor=magenta,   
}

\begin{document}

\maketitle


\begin{abstract}

  \renewcommand{\thefootnote}{*}
  \footnotetext{Corresponding author}
  \renewcommand{\thefootnote}{\arabic{footnote}}
  
  \noindent In recent years, emotional Text-to-Speech (TTS) synthesis and emphasis-controllable speech synthesis have advanced significantly. However, their interaction remains underexplored. We propose Emphasis Meets Emotion TTS (EME-TTS), a novel framework designed to address two key research questions: (1) how to effectively utilize emphasis to enhance the expressiveness of emotional speech, and (2) how to maintain the perceptual clarity and stability of target emphasis across different emotions. EME-TTS employs weakly supervised learning with emphasis pseudo-labels and variance-based emphasis features. Additionally, the proposed Emphasis Perception Enhancement (EPE) block enhances the interaction between emotional signals and emphasis positions. Experimental results show that EME-TTS, when combined with large language models for emphasis position prediction, enables more natural emotional speech synthesis while preserving stable and distinguishable target emphasis across emotions. Synthesized samples are available on-line\footnote{\url{https://wd-233.github.io/EME-TTS_DEMO/}}. 
  

\end{abstract}

\section{Introduction}

With the advancement of deep learning, Text-to-Speech (TTS) systems have significantly improved in terms of quality, clarity, and naturalness, leveraging architectures such as Transformers \cite{waswani2017attention,ren2020fastspeech}, normalizing flows \cite{kim2020glow,kim2021conditional}, and diffusion models \cite{popov2021grad,tan2024naturalspeech}. However, conventional TTS systems often struggle with expressiveness, producing monotonal and mechanical speech. To address this issue, researchers have explored emotionally expressive speech synthesis \cite{ExpressiveFastSpeech2}, employing explicit emotion labels \cite{tits2020exploring,diatlova23_ssw,guo2023emodiff} and reference-based approaches \cite{li2021controllable,lei2022msemotts,zaidi22b_interspeech} to enhance speech expressiveness.

A key aspect of expressive speech is emphasis, which highlights prominent prosodic regions through variations in pitch, phoneme duration, and spectral energy \cite{eriksson2015acoustics,eriksson2016acoustics,eriksson2020lexical}. Several studies have introduced emphasis control in TTS \cite{li2018emphatic,shechtman2021supervised,seshadri22_interspeech}, ranging from handcrafted feature integration in systems based on Hidden Markov Models (HMM) \cite{li2018emphatic} to leveraging intermediate acoustic cues like pitch range \cite{shechtman2021supervised} and variance-based features \cite{seshadri22_interspeech}.

Despite significant progress in emotional TTS and emphasis-controllable synthesis, the interplay between emotion and emphasis in speech remains largely unexplored. Emphasis and emotion are intrinsically linked—emphasis modulates emotional perception by influencing prosodic patterns, while emotional states naturally determine which words are accentuated in speech. This study develops a speech synthesis model capable of simultaneously controlling both emotion and emphasis. Specifically, we seek to answer two key questions: How can emphasis be effectively leveraged to enhance the expressiveness of emotional speech? And how can emphasis clarity and stability be preserved across different emotional conditions?

To answer the first question, we examine how predefined emphasis positions influence emotional speech synthesis. Emotional speech is shaped by both prosody and semantics, where shifts in emphasis position and intensity can alter an utterance’s emotional interpretation. While prior works such as EE-TTS \cite{zhong23_interspeech} predicts emphasis using textual and grammatical cues, we argue that Large Language Models (LLMs) \cite{zhao2023survey} can infer emphasis more effectively. Instead of focusing on semantic-based emphasis prediction, we investigate how predefined emphasis positions influence emotional speech synthesis. To this end, we annotate emphasis pseudo-labels on an emotional speech dataset and integrate an improved variance-based emphasis modeling approach. During inference, an LLM predicts emphasis positions based on emotion labels and text input. Evaluations show that our model generates more emotionally expressive speech, especially when contextual information is present.

To answer the second question, we introduce the Emphasis Perception Enhancement (EPE) block to improve emphasis clarity and stability across emotions. This block refines the interaction between emotion control signals and emphasis locations, reducing unintended interference from global emotional effects. Additionally, this design mitigates artifacts that often arise at emphasized positions, enhancing both perceptual emphasis clarity and synthesis quality. 

To summarize, the main contributions of this work are as follows:
\begin{itemize}
\item Emotionally Controllable Emphasis Modeling: We incorporate variance-based emphasis features into an emotional TTS framework, and use weakly supervised learning with EmphaClass \cite{seyssel-etal-2024-emphassess} pseudo-labeling on the ESD dataset \cite{zhou2022emotional}.

\item Refined Emotion-Emphasis Interaction: We propose EPE block to modulate emphasis prominence effectively, which ensures stable and clear emphasis across emotions.

\item To the best of our knowledge, this study is the first to systematically investigate the relationship between emotion and emphasis in speech synthesis. EME-TTS enhances emotional expressiveness and preserves emphasis clarity with predefined emphasis positions and emotion labels.
\end{itemize}

\begin{figure*}[htbp]
    \setlength{\abovecaptionskip}{5pt}  
    \setlength{\belowcaptionskip}{5pt}  
    \centering
    \begin{minipage}[t]{0.22\linewidth}  
        \centering
        \mbox{\adjustbox{valign=m}{\includegraphics[width=1.5in]{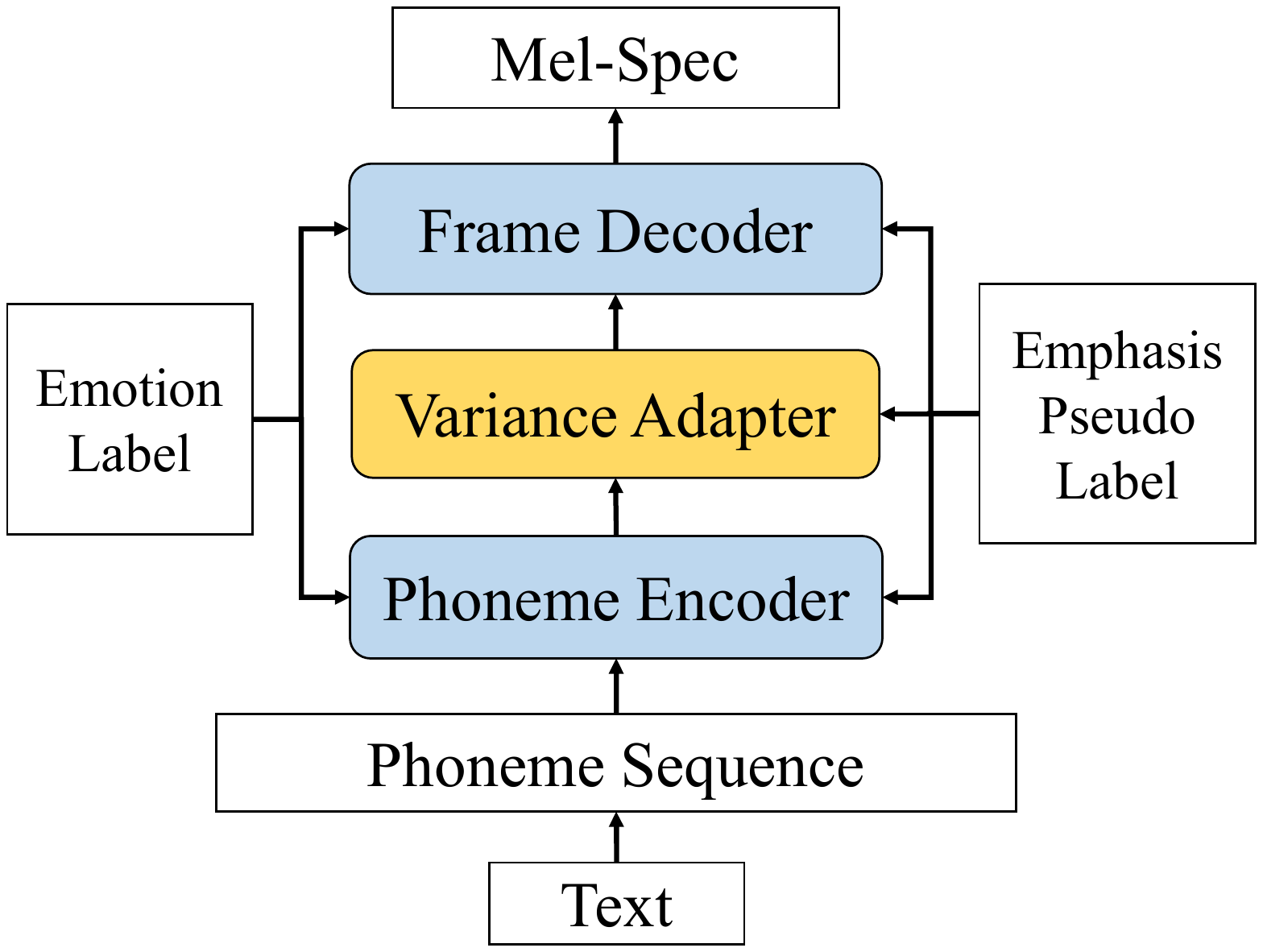}}}  
        \vspace{28pt} \\  
        \subcaption{Overall Architecture}\label{figure4}  
    \end{minipage}
    \begin{minipage}[t]{0.43\linewidth}  
        \centering
        \mbox{\adjustbox{valign=m}{\includegraphics[width=3.0in]{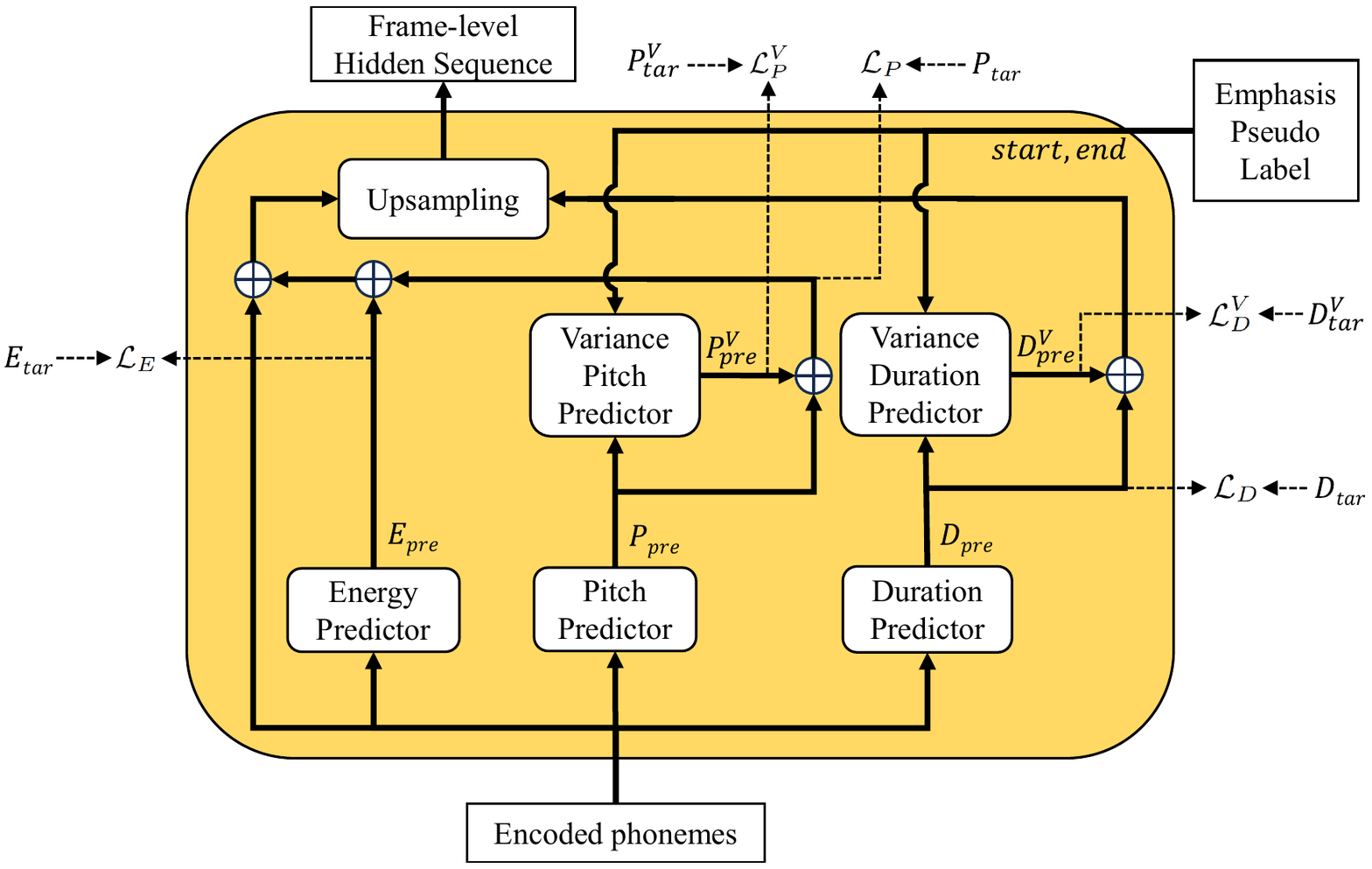}}}  
        \vspace{-12pt} \\  
        \subcaption{Variance Adapter}\label{figure2}  
    \end{minipage}
    \begin{minipage}[t]{0.34\linewidth}  
        \centering
        \mbox{\adjustbox{valign=m}{\includegraphics[width=2.4in]{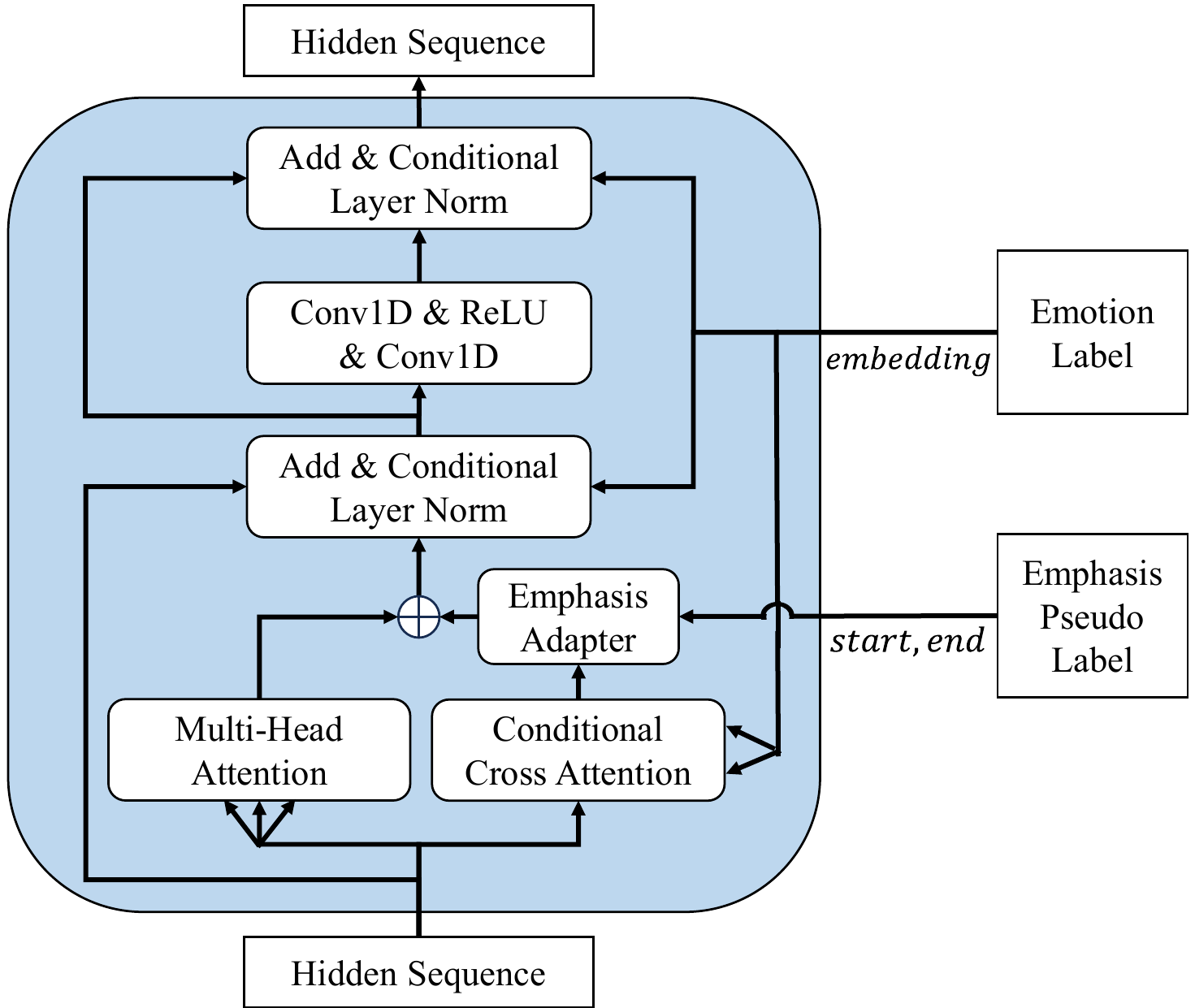}}}  
        \vspace{1pt} \\  
        \subcaption{EPE Block}\label{figure3}  
    \end{minipage}
    \caption{The entire framework of our proposed model. (a) is the overall architecture diagram. (b) and (c) show the detailed structure of variance adapter and Emphasis Perception Enhancement (EPE) block, respectively.} 
    \label{figure1}
    \vspace{-16pt}
\end{figure*}

\section{Proposed Method}

\subsection{Overview}

The overall architecture of EME-TTS is shown in Figure 1a. We use EmoSpeech \cite{diatlova23_ssw} as the base architecture of the acoustic model, using the embedding of emotions and the position of the emphasis as conditions. These conditions are obtained through the emotion label and the emphasis pseudo label respectively. EME-TTS consists of a phoneme encoder, a variance adapter, and a frame decoder, working in conjunction to generate emotionally expressive speech with controllable emphasis. The phoneme encoder processes the input phoneme sequence extracted from text, encoding phonetic features that serve as the foundation for speech synthesis. The variance adapter models prosodic variations through duration, pitch, and energy while integrating variance-based emphasis features for explicit emphasis control. The resulting frame-level hidden representations are then converted into a mel-spectrogram by the frame decoder. 

\subsection{Weakly Supervised Emphasis Pseudo-Labeling}

Determining whether a word is emphasized within a sentence involves significant subjectivity, as multiple valid emphasis positions may exist. Consequently, collecting large-scale labeled emphasis data is both challenging and costly. Unlike EE-TTS \cite{zhong23_interspeech}, which employs a wavelet-based prosody toolset \cite{suni2017hierarchical} to compute prominence scores via Continuous Wavelet Transform (CWT) using pitch, energy, and duration signals, we leverage the EmphAssess dataset and the EmphaClass emphasis recognizer \cite{seyssel-etal-2024-emphassess}. EmphaClass fine-tunes a pre-trained Self-Supervised Learning (SSL) speech model for frame-level classification, and aggregates these scores to determine word-level emphasis with high accuracy. By utilizing EmphaClass to annotate emphasis in the ESD dataset \cite{zhou2022emotional}, which underpins our TTS experiments, we obtain highly reliable emphasis pseudo-labels without the need for extensive manual annotation.

\subsection{Variance Adapter}

In order to integrate both fundamental prosodic predictors and variance-based emphasis modeling, the variance adapter is designed to model and regulate prosodic variations. As shown in Figure 1b, it consists of predictors for pitch, duration, and energy, which predict their respective prosodic features from encoded phonemes. Additionally, it includes variance-based pitch and duration predictors, which refine these predictions by capturing local deviations in emphasized regions. The upsampling mechanism ensures that phoneme-level prosodic features are aligned with the frame-level hidden sequence before being passed to the decoder.

To explicitly control emphasis, we incorporate variance-based prosodic features as local modulation signals. Following the assumption in \cite{seshadri22_interspeech} that pitch and duration are the primary indicators of emphasis, with energy having a lesser impact, we focus on modeling only pitch and duration variance features, omitting energy. The computation of these variance features is formulated as follows:
\begin{equation}
\text{Pitch Variance} = W_{F_0} - S_{F_0}
\end{equation}
\begin{equation}
\text{Duration Variance} = W_{\text{dur}} - S_{\text{dur}}
\end{equation}
where \( W_{F_0} \) and \( W_{\text{dur}} \) represent the average pitch and duration of phonemes in emphasized regions, while \( S_{F_0} \) and \( S_{\text{dur}} \) denote the sentence-level averages. These values are derived within the Variance Pitch Predictor and Variance Duration Predictor using the emphasis pseudo-labels, which provide the ${start, end}$ positions of emphasized words. 

During training, the predicted pitch feature \( P_{\text{pre}} \) is used to compute the pitch variance feature \( P^V_{\text{pre}} \) via Equation (1), and its loss is calculated against the target pitch feature \( P_{\text{tar}} \):
\begin{equation}
\mathcal{L}_{P} = \text{MSE}(P_{\text{pre}} + P^V_{\text{pre}}, P_{\text{tar}})
\end{equation}
where Mean Squared Error (MSE) measures the average squared difference between predicted and target values. Similarly, the pitch variance loss is:
\begin{equation}
\mathcal{L}_{P}^{V} = \text{MSE}(P^V_{\text{pre}}, P^V_{\text{tar}})
\end{equation}
For duration modeling, we adopt the same approach:
\begin{equation}
\mathcal{L}_{D} = \text{MSE}(D_{\text{pre}} + D^V_{\text{pre}}, D_{\text{tar}})
\end{equation}
\begin{equation}
\mathcal{L}_{D}^{V} = \text{MSE}(D^V_{\text{pre}}, D^V_{\text{tar}})
\end{equation}
where the variance features \( P^V_{\text{pre}} \) and \( D^V_{\text{pre}} \) are only applied to emphasized regions, with non-emphasized regions set to zero. Since direct variance calculations may introduce extreme values or negative emphasis scores, we normalize the features to the range of [0,2] based on data distribution and apply regularization for stability. The aforementioned losses, along with the energy loss, are included in the final total loss calculation.

\subsection{Emphasis Perception Enhancement Block}

The phoneme encoder and frame decoder in EME-TTS are composed of multiple stacked Emphasis Perception Enhancement (EPE) blocks, replacing the original feed-forward transformer blocks, as illustrated in Figure 1c. Each EPE block refines the modeling of emphasis perception by integrating Multi-Head Attention (MHA), Conditional Cross Attention (CCA), and Emphasis Adapter (EA), ensuring that the synthesized speech maintains clear and stable emphasis across emotions. The input hidden sequence is processed through self-attention and conditional normalization layers before being further refined by convolutional layers. Simultaneously, the emotion embedding and emphasis position can serve as external conditions.

Among these components, MHA captures global dependencies within the hidden sequence by computing multiple attention heads, allowing the model to extract contextual relationships between phonemes. CCA re-weights self-attention by incorporating emotional cues, allowing attention distributions to be adjusted based on the given emotion embedding $c$. CCA re-weights self-attention using:
\begin{equation}
Q = W_q \cdot h, \quad K = W_k \cdot c, \quad V = W_v \cdot c
\end{equation}
\begin{equation}
w = \text{softmax} \left( \frac{Q \cdot K^T}{\sqrt{d}} \right)
\end{equation}
\begin{equation}
\text{CCA} = w \cdot V
\end{equation}
where \( h \) represents the input feature, and \( w \) denotes computed attention weights. In expressive speech, the prominence of certain words naturally varies with emotion, causing inconsistencies in emphasis perception. This can lead to unintended shifts in emphasis positions or even the suppression of intended emphasis under strong emotional conditions. Additionally, increasing the prominence of emphasized words may introduce artifacts and degrade synthesis quality.

To mitigate these issues, we introduce EA that refines attention distributions in predefined emphasis regions. EA explicitly enhances emphasis perceptibility while minimizing interference from emotional variations. Given an initial attention weight \( w \), the adjusted weight is computed as:
\begin{equation}
w_{\text{adjusted}} = w + \Delta w, \quad \Delta w = {strength} \cdot {mask}({start, end})
\end{equation}
where \({mask(start, end)}\) identifies the designated emphasis positions, and ${strength}$ scales the emphasis intensity. This enhancement serves two key purposes: first, it ensures that emphasized words remain perceptually distinct, even in highly expressive speech; second, it refines emphasis representation through attention modulation rather than direct energy or pitch manipulation, thereby reducing synthesis artifacts.

For Conditional Layer Normalization (CLN), we adopt a design similar to AdaSpeech4 \cite{wu22f_interspeech} to integrate emotional context into the normalization process, ensuring adaptive prosodic control across different emotional conditions.

\section{Experiments}

\subsection{Experimental Setup}
For our experiments, EME-TTS utilizes the English portion of the Emotional Speech Database (ESD) \cite{zhou2022emotional}, which comprises recordings from 10 speakers across five emotions: \textit{angry, happy, sad, surprise, and neutral}. Each speaker contributes 350 utterances per emotion, resulting in approximately 1,750 utterances and 1.2 hours of speech per speaker. We follow the train/validation/test splits established in EmoSpeech \cite{diatlova23_ssw}, where the validation and test sets consist of 19 and 31 utterances per emotion per speaker, leading to a total of 950 and 1,550 utterances, respectively. Emphasis positions in the training data are labeled using EmphaClass \cite{seyssel-etal-2024-emphassess}. 

The model utilizes iSTFTNet \cite{kaneko2022istftnet} as the vocoder, which is trained on the English subset of the ESD dataset \cite{zhou2022emotional}. The strength in EA is set to 0.2. Training is conducted on 2 Nvidia A100 GPUs and 8 RTX 4090 GPUs, with a batch size of 64 for 100,000 steps. The Adam optimizer \cite{diederik2014adam} is used with a learning rate of 0.0001 and ($\beta_1$, $\beta_2$) = (0.5, 0.9). 


\subsection{Evaluation Metrics}

The evaluation framework comprises both objective and subjective assessments. The objective evaluation includes audio quality assessment and emotional accuracy measurement. The subjective evaluation consists of four tasks: 
\begin{itemize}
\item Emphasis Accuracy Test (EAT): Measures how well predicted emphasis positions match listener perception.

\item Emotion Accuracy Test (EAT-EMO): Assesses emotion recognition accuracy by comparing perceived and intended emotions.

\item Emotional Expressiveness Preference Test (EEPT): Listeners select the most emotionally expressive sample from multiple outputs, with preference scores indicating expressiveness strength.

\item Mean Opinion Score (MOS) Rating: Evaluates naturalness and quality on a five-point scale.
\end{itemize}

A total of 11 listeners participated in all evaluations, with each participant completing all four tasks. For all tasks, higher scores indicate better performance. 

\subsubsection{Emphasis Accuracy}

To demonstrate EME-TTS's ability to preserve the perceptual clarity and stability of target emphasis across different emotions, we designed a more challenging subjective test as task 1 (EAT), distinct from previous studies on emphasis control \cite{seshadri22_interspeech}. Instead of rating the degree of emphasis at a predefined position, participants were asked to identify the emphasized words in 80 randomly shuffled speech samples. These samples were generated from texts in the test set using EME-TTS w/o EPE and EME-TTS. The results presented in Table 1 indicate that the emphasis produced by our proposed model was clearly perceivable across different emotions.

Notably, compared to other emotions, the \textit{surprise} emotion posed a greater challenge for listeners in accurately identifying the emphasis position. This is attributed to the frequent pitch rise at the end of \textit{surprise} speech, which often led participants to mistakenly perceive the emphasized word as being at the sentence’s end. The integration of the EPE effectively mitigated this issue by enhancing the prominence of the intended emphasis position. At the same time, it improved the distinctiveness of emphasis recognition across all emotions.

\begin{table}[ht]
\caption{Emphasis Recognition Accuracy of Different Models.}
\label{tab:emphasis_recognition}
\centering
\renewcommand{\arraystretch}{1}  
\resizebox{0.48\textwidth}{!}{  
\fontsize{18pt}{24pt}\selectfont  
\begin{tabular}{ l c c c c c c }
\toprule
\textbf{Model} & \textbf{Mean} & \textbf{Neutral} & \textbf{Angry} & \textbf{Happy} & \textbf{Sad} & \textbf{Surprise} \\
\midrule
EME-TTS w/o EPE & 0.73 & 0.77 & 0.75 & \textbf{0.82} & \textbf{0.75} & 0.55 \\
\midrule
\textbf{EME-TTS} & \textbf{0.78} & \textbf{0.80} & \textbf{0.82} &\textbf{0.82} & \textbf{0.75} &\textbf{0.64} \\
\bottomrule
\end{tabular}
}
\vspace{-8pt}
\end{table}

\subsubsection{Emotion Accuracy}

To demonstrate that controlling emphasis in EME-TTS enhances the accuracy of perceived emotions in synthesized speech, we conducted both objective and subjective evaluations. For comparison, we included EmoSpeech \cite{diatlova23_ssw} and CosyVoice2-0.5B-Instruct (CosyVoice2) \cite{du2024cosyvoice} as baseline models. CosyVoice2 provides multiple inference modes; To ensure fairness, all CosyVoice2-generated samples were conditioned on a \textit{neutral} reference speaker’s audio and a textual emotion prompt, ensuring that only the speaker's identity and emotion labels were provided as input. During inference, our proposed model consistently utilized a large language model \cite{achiam2023gpt} to predict suitable emphasis positions, which were then used as input for testing.

\begin{table}[ht]
\vspace{-4pt}
\caption{Objective Evaluation of Different Models on Emotion Accuracy.  }
\label{tab:objective_evaluation}
\centering

\resizebox{0.46\textwidth}{!}{  
\fontsize{18pt}{24pt}\selectfont  
\begin{tabular}{ l c c c c c c }
\toprule
\textbf{Model} & \textbf{Mean} & \textbf{Neutral} & \textbf{Angry} & \textbf{Happy} & \textbf{Sad} & \textbf{Surprise} \\
\midrule

CosyVoice2 \cite{du2024cosyvoice}       & 0.68 & \textbf{0.99}  & 0.51 & \textbf{0.73} &  0.52&0.36 \\
EmoSpeech \cite{diatlova23_ssw}        & 0.72 & \textbf{0.99} & \textbf{0.91} & 0.69 & 0.54 & \textbf{0.48} \\
EME-TTS w/o EPE & \textbf{0.74} & \textbf{0.99} & 0.90 & 0.71 & 0.60 & 0.47 \\
\midrule
\textbf{EME-TTS}           & 0.73 & \textbf{0.99} & 0.87 & 0.70 & \textbf{0.61} & 0.47 \\
\bottomrule
\end{tabular}
}
\vspace{-4pt}
\end{table}

From an objective perspective, we utilized the Emotion2vec-plus-large model \cite{ma-etal-2024-emotion2vec} to evaluate the emotional accuracy of 1,550 synthesized audio samples from each model in the test set. The recognition outcome was assigned a score of 1 for correct classifications and 0 for incorrect ones, from which the overall accuracy was computed. The results shown in Table 2 indicate that while local emphasis control did not introduce substantial changes in global prosody, it notably improved the recognition of \textit{sad} emotions in objective evaluation. This improvement is attributed to the increased duration of emphasized regions, which effectively enhanced emotion perception.

On the subjective side, in task 2 (EAT-EMO), participants were asked to identify the emotions of 80 randomly shuffled audio samples. Similarly, scores of 1 and 0 were assigned for correct and incorrect classifications respectively, to assess the emotional accuracy of the synthesized speech. However, as shown in Table 3, the subjective evaluation revealed unexpected shifts in emotion perception due to the introduction of emphasis. First, emphasis increased the likelihood of \textit{neutral} speech being perceived as emotional, leading to a slight decrease in the accuracy of \textit{neutral} emotion expression. Second, unlike CosyVoice2 \cite{du2024cosyvoice}, which struggled to synthesize \textit{angry} and \textit{sad} emotions in this inference mode, EME-TTS produced speech that made these emotions more perceptible, outperforming both CosyVoice2 \cite{du2024cosyvoice} and EmoSpeech \cite{diatlova23_ssw}. For \textit{happy} and \textit{surprise} emotions, EME-TTS achieved accuracy levels comparable to other models. These results demonstrate that EME-TTS achieves higher emotion accuracy in synthesized speech compared to baseline models, highlighting its overall effectiveness in generating emotionally expressive speech.

\begin{table}[ht]

\vspace{-1pt}
\caption{Subjective Evaluation of Different Models on Emotion Accuracy.  }
\label{tab:subjective_evaluation}
\centering
\resizebox{0.46\textwidth}{!}{  
\fontsize{18pt}{24pt}\selectfont  
\begin{tabular}{ l c c c c c c }
\toprule
\textbf{Model} & \textbf{Mean} & \textbf{Neutral} & \textbf{Angry} & \textbf{Happy} & \textbf{Sad} & \textbf{Surprise} \\

\midrule

CosyVoice2 \cite{du2024cosyvoice}       & 0.48 & \textbf{0.93} & 0.07 & 0.34 & 0.36 & \textbf{0.70} \\
 EmoSpeech \cite{diatlova23_ssw}        & 0.58 & 0.86 & 0.48 & \textbf{0.36} & 0.50 & \textbf{0.70} \\
 EME-TTS w/o EPE & 0.58 & 0.68 & 0.57 & 0.27 & 0.80 & 0.59 \\
\midrule
\textbf{EME-TTS}           & \textbf{0.67} & 0.80 & \textbf{0.75} & 0.32 & \textbf{0.82} & 0.68 \\
\bottomrule
\end{tabular}
}
\vspace{-8pt}
\end{table}

\subsubsection{Improvement of Emotional Expressiveness Through Emphasis}

To assess how emphasis enhances emotional expressiveness in EME-TTS, we conducted a ranking experiment as task 3 (EEPT). Participants evaluated 30 sets of speech samples, each containing outputs from four models, based on perceived emotional expressiveness. Among them, 10 sets were derived from a short passage with contextual information. Within each set, samples were ranked from 1 (least expressive) to 4 (most expressive). Figure 2 illustrates the ranking distribution of emotional expressiveness across different TTS models. Participants evaluated speech samples based on perceived emotional expressiveness, with (a) assessing individual sentences and (b) ranking sentences within a contextualized passage. The results indicate that EME-TTS consistently received the highest rankings, especially in the contextualized setting. This suggests that surrounding linguistic context strengthens the semantic foundation for emphasis, further enhancing emotional expressiveness.

\begin{figure}[htbp]
    \vspace{-7pt}
    \setlength{\abovecaptionskip}{5pt}  
    \setlength{\belowcaptionskip}{5pt}  
    \centering
    \begin{minipage}[t]{0.5\linewidth}
        \centering
        \includegraphics[width=\linewidth]{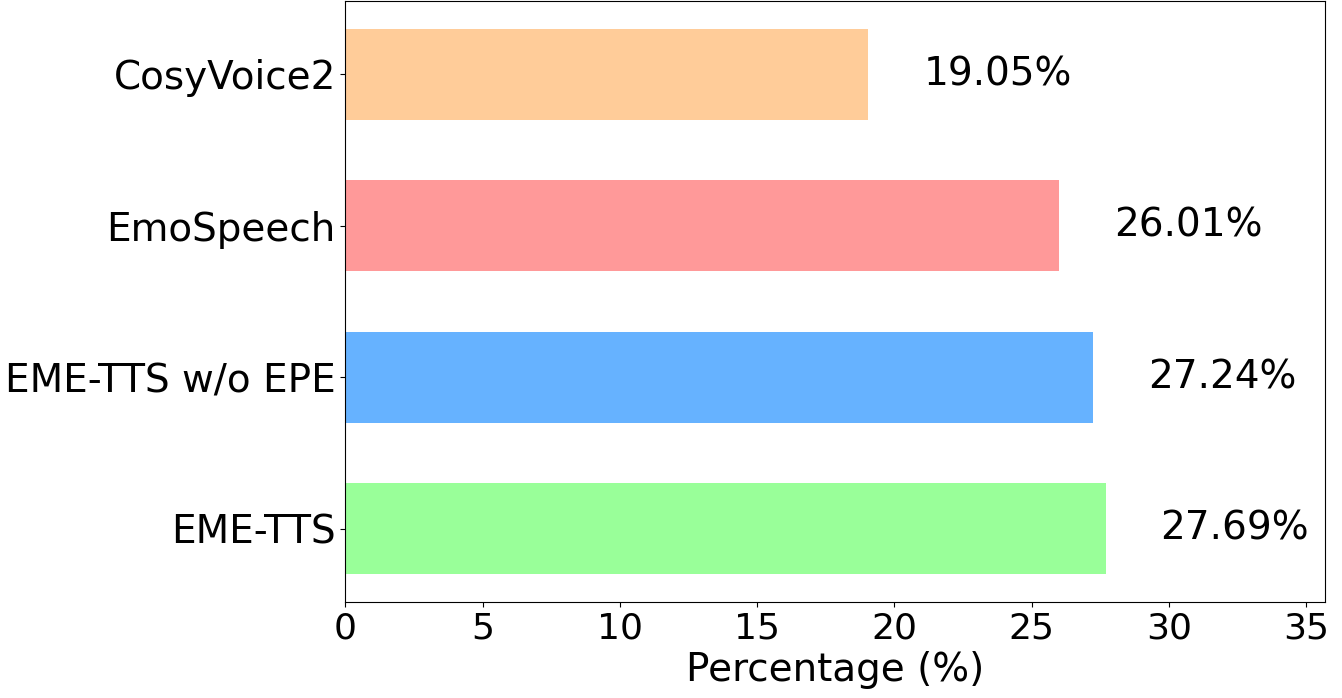}
        \subcaption{Isolated Sentences}\label{figure4}  
    \end{minipage}%
    \begin{minipage}[t]{0.5\linewidth}
        \centering
        \includegraphics[width=\linewidth]{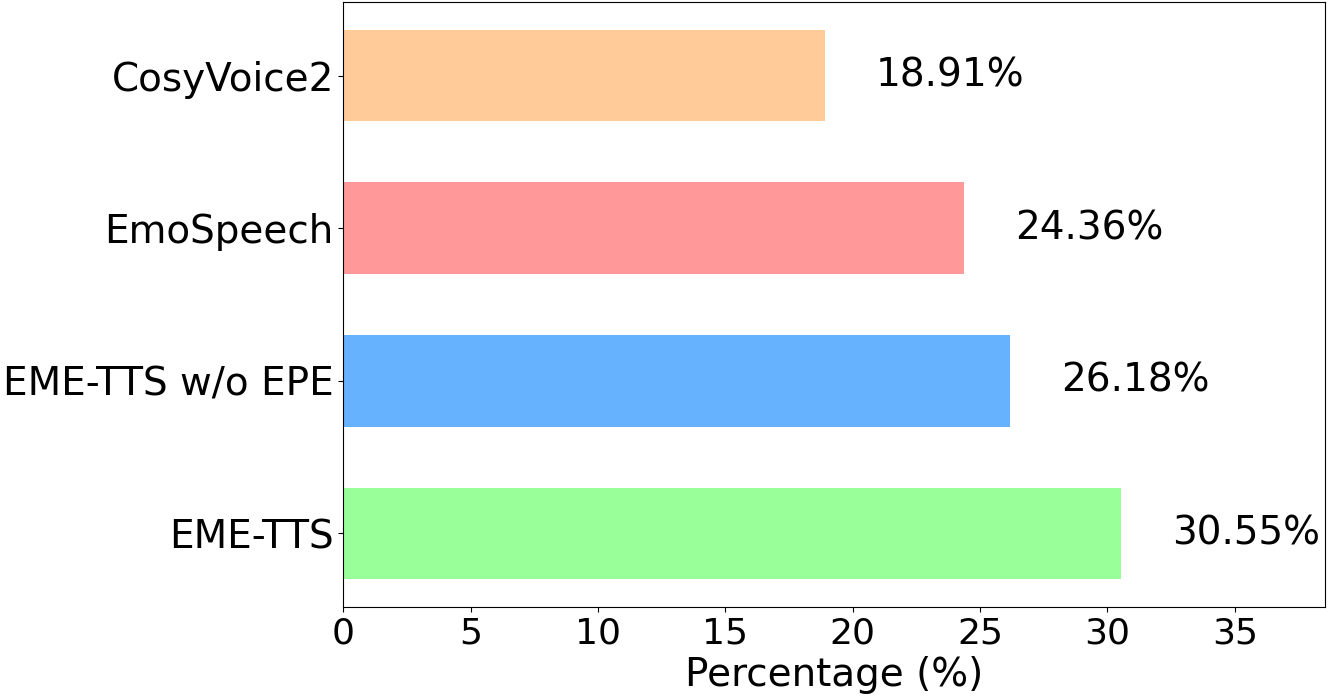}
        \subcaption{Contextualized Sentences}\label{figure2}  
    \end{minipage}
    \caption{Effect of Emphasis on Emotional Expressiveness in TTS Models.} 
    \label{figure1}
    \vspace{-20pt}
\end{figure}

\subsubsection{Speech Quality and Naturalness}

We assess the quality and naturalness of the synthesized speech through both objective and subjective evaluations. Objectively, we utilize the NISQA library \cite{mittag20_interspeech} to predict naturalness scores ratings on a 5-point scale. Subjectively, participants completed task 4 (MOS Rating), in which they rated 100 randomly shuffled speech samples for overall audio quality and naturalness, ranging from 1 (bad) to 5 (excellent). These 100 utterances were selected from the test set, ensuring an equal distribution of emotions. Table 4 indicates that EME-TTS mitigates artifacts introduced by emphasis control,  enhancing synthesis quality.

\begin{table}[ht]
\vspace{-2pt}
\caption{Comparison of Models for MOS and NISQA Scores. }
\label{tab:model_comparison}
\centering
\resizebox{0.350\textwidth}{!}{  
\begin{tabular}{ l c c }
\toprule
\textbf{Model} & \textbf{MOS (\( \uparrow \))} & \textbf{NISQA (\( \uparrow \))} \\

\midrule

Original                   & 4.94 $\pm$ 0.03  & 4.17 $\pm$ 0.57 \\
Reconstructed              & 4.86 $\pm$ 0.08 & 4.11 $\pm$ 0.58 \\
\midrule
EmoSpeech \cite{diatlova23_ssw}                 & 4.14 $\pm$ 0.20 & 3.71 $\pm$ 0.74 \\
EME-TTS w/o EPE            & 3.98 $\pm$ 0.32 & 3.66 $\pm$ 0.68 \\
\midrule
\textbf{EME-TTS}                 &  \textbf{4.22 $\pm$ 0.28} & \textbf{3.76 $\pm$ 0.60} \\

\bottomrule
\end{tabular}
}
\vspace{-14pt}
\end{table}

\section{Conclusion}
This paper presents EME-TTS, a framework that explores how emphasis enhances emotional expressiveness and how to maintain its perceptual clarity and stability across emotions. By leveraging variance-based emphasis features, weakly supervised learning, and EPE, EME-TTS demonstrate its effectiveness in generating emotionally expressive speech with clear and controllable emphasis. For future work, we aim to further investigate the role of different emphasis strategies in enhancing emotional expressiveness, particularly in achieving improvements for specific emotions where the current model's enhancements remain limited.

\section{Acknowledgements}
This work was supported in part by the Scientific Research Staring Foundation of Hangzhou Institute for Advanced Study (2024HIASC2001), in part by Zhejiang Provincial Natural Science Foundation of China (No. LQN25F020001), and in part by the Key R\&D Program of Zhejiang  (2025C01104).

\bibliographystyle{IEEEtran}
\bibliography{mybib}

\end{document}